\documentclass[preprint,showpacs,preprintnumbers,superscriptaddress]{revtex4}
\usepackage{amssymb}
\usepackage{graphicx}
\usepackage{amsmath}
\usepackage{bm}
\usepackage{dcolumn}
\usepackage{amsfonts}
\usepackage{color}

\begin{document}

\title{Negative Off-Diagonal Conductivities in a Weakly Coupled Quark Gluon
Plasma}
\author{Jiunn-Wei Chen}
\email{jwc@phys.ntu.edu.tw}
\affiliation{Department of Physics, National Center for Theoretical Sciences, and Leung
Center for Cosmology and Particle Astrophysics, National Taiwan University,
Taipei 10617, Taiwan}
\author{Yen-Fu Liu}
\email{liuyenfu@gmail.com}
\affiliation{Department of Physics, National Center for Theoretical Sciences, and Leung
Center for Cosmology and Particle Astrophysics, National Taiwan University,
Taipei 10617, Taiwan}
\author{Shi Pu}
\email{pushi@ntu.edu.tw}
\affiliation{Department of Physics, National Center for Theoretical Sciences, and Leung
Center for Cosmology and Particle Astrophysics, National Taiwan University,
Taipei 10617, Taiwan}
\affiliation{Interdisciplinary Center for Theoretical Study and Department of Modern
Physics,University of Science and Technology of China, Anhui 230026,
People's Republic of China}
\author{Yu-Kun Song}
\email{songyk@ustc.edu.cn}
\affiliation{Interdisciplinary Center for Theoretical Study and Department of Modern
Physics,University of Science and Technology of China, Anhui 230026,
People's Republic of China}
\affiliation{Key Laboratory of Quark and Lepton Physics (Central China Normal
University), Ministry of Education, People's Republic of China}
\author{Qun Wang}
\email{qunwang@ustc.edu.cn}
\affiliation{Interdisciplinary Center for Theoretical Study and Department of Modern
Physics,University of Science and Technology of China, Anhui 230026,
People's Republic of China}

\begin{abstract}
We calculate the conductivity matrix of a weakly coupled quark-gluon plasma
at the leading-log order. By setting all quark chemical potentials to be
identical, the diagonal conductivities become degenerate and positive, while
the off-diagonal ones become degenerate but negative (or zero when the
chemical potential vanishes). This means a potential gradient of a certain
fermion flavor can drive backward currents of other flavors. A simple
explanation is provided for this seemingly counter intuitive phenomenon. It
is speculated that this phenomenon is generic and most easily measured in
cold atom experiments.
\end{abstract}

\pacs{25.75.Nq, 12.38.Mh}
\maketitle

\section{Introduction}

Hydrodynamics describes the evolution of a fluid perturbed away from thermal
equilibrium by long wave length fluctuations. The long wave length physics
(long compared with the mean field path of particle collisions) can be
systematically described by an expansion of space-time derivatives on
classical fields with prefactors called transport coefficients. These
transport coefficients encode the physics of short (compared with the mean
free path) distance and are inputs to hydrodynamics. But they can be
computed, in principle, once the microscopic theory of the system is known.

We are interested in computing the transport coefficients in Quantum
Chromodynamics (QCD) with $N_{f}$ flavors of massless quarks at finite
temperature ($T$) and chemical potentials ($\mu _{a}$, $a=1,2,\cdots ,N_{f}$%
). The leading transport coefficients at the first derivative order include
the shear viscosity ($\eta $), bulk viscosity ($\zeta $), and the
conductivity matrix ($\lambda $).

The shear viscosity of QCD has attracted a lot of attention recently. Its
ratio with the entropy density ($s$) extracted from the hot and dense matter
created at Relativistic Heavy Ion Collider (RHIC) \cite%
{Arsene:2004fa,Adcox:2004mh,Back:2004je,Adams:2005dq} just above the phase
transition temperature ($T_{c}$) yields $1/(4\pi )\leq \eta /s\leq 2.5/(4\pi
)$ at $T_{c}\leq T\leq 2T_{c}$ \cite{Song:2010mg}, which is close to a
conjectured universal lower bound of $1/(4\pi )$ \cite{Kovtun:2004de}
inspired by the gauge/gravity duality \cite%
{Maldacena:1997re,Gubser:1998bc,Witten:1998qj}. This value of $\eta /s$
cannot be explained by extrapolating perturbative QCD result \cite%
{Arnold:2000dr,Arnold:2003zc,Chen:2010xk,Chen:2011km}. The smallest $\eta /s$
is likely to exist near $T_{c}$ \cite{Csernai:2006zz,Chen:2006iga} (see,
e.g., Ref. \cite{Chen:2011km} for a compilation and more references). Also\
finite $\mu $ results suggests that $\eta /s$ is smaller at smaller $\mu $.
This is based on results of perturbative QCD at $T\gg $ $T_{c}$ \cite%
{Chen:2012jc} and of a hadronic gas at $T\ll $ $T_{c}$ and small $\mu $ \cite%
{Chen:2007xe}. It is speculated that the same pattern will persist at $T_{c}$
such that the smallest $\eta /s$ might exist near $T_{c}$ with $\mu =0$ \cite%
{Chen:2012jc}.

For the bulk viscosity, the sum rule study \cite%
{Kharzeev:2007wb,Karsch:2007jc} shows that $\zeta $ increases rapidly near $%
T_{c}$ when $T$ approaches $T_{c}$ from above. This is consistent with the
lattice gluon plasma result near $T_{c}$ \cite{Meyer:2010ii} and
perturbative QCD result \cite{Arnold:2006fz} at much higher $T$. This, when
combined with pion gas results below $T_{c}$ \cite%
{Chen:2007kx,FernandezFraile:2008vu,Lu:2011df,Dobado:2011qu,Chakraborty:2010fr}%
, suggests that $\zeta /s$ has a local maximum near $T_{c}$ (see, e.g., \cite%
{Chen:2011km} for a compilation). Unlike $\eta /s$, perturbative QCD result
shows very small $\mu $ dependence in $\zeta /s$\ \cite{Chen:2012jc}. Note
that at high $\mu $, there are also bulk viscosities governed by the weak
interaction such as the Urca processes which have consequences in neutron
star physics \cite%
{Dong:2007mb,Alford:2006gy,Alford:2008pb,Sa'd:2006qv,Sa'd:2007ud,Wang:2010ydb}%
. These are quite different from the transport coefficients from the strong
interaction mentioned above.

The perturbative QCD calculations of $\eta $ and $\zeta $ with finite $\mu $
were performed at the leading-log (LL) order of the strong coupling constant
($g$) expansion in Ref. \cite{Chen:2012jc}. Either $T$ or $\mu $ in the
calculation is much larger than $\Lambda _{QCD}$ which is the scale where
QCD becomes non-perturbative. But the calculation is not applicable to the
color superconducting phase at $\mu /T\rightarrow \infty $, since the vacuum
in the calculation has no symmetry breaking.

In this work, we apply the same perturbative QCD approach to compute the
conductivity matrix $\lambda $ at the LL order. The conductivity is an
important transport coefficient which plays an essential role in the
evolution of electromagnetic fields in heavy ion collisions \cite%
{Huang:2013iia,McLerran:2013hla}. The conductivity in strongly coupled quark
gluon plasma was calculated with lattice QCD \cite{Ding:2010ga,Amato:2013naa}
and Dyson-Schwinger equation \cite{Qin:2013aaa}.

We first review the constraints from the second law of thermal dynamics
(i.e. the entropy production should be non-negative) which show that the
particle diffusion, heat conductivities, and electric conductivity are all
unified into one single conductivity in this system. When $N_{f}>1$, the
conductivity becomes a $N_{f}\times N_{f}$ matrix. We then show through the
Boltzmann equation that the conductivity matrix $\lambda $ at the LL order
is symmetric and positive definite ($\sum_{a,b}\lambda _{ab}X^{a}X^{b}>0$
for any real, non-vanishing vector $X$). The former is a manifestation of
the Onsager relation while the latter is a manifestation of the second law
of thermal dynamics.

For simplicity, we show the numerical results of $\lambda $ with all fermion
chemical potential to be identical. In this limit, there are only two
independent entries in $\lambda $. All the diagonal matrix elements are
degenerate and positive since $\lambda $ is positive definite. However, the
off-diagonal matrix elements are degenerate but negative at finite $\mu $.
This means a gradient $\bm{\nabla }\mu _{a}$ can drive a current of flavor $%
a $ alone the gradient direction, but it will also drive currents of
different flavors in the opposite direction. This backward current
phenomenon might seem counter intuitive, but we find that it is generic and
it has a simple explanation. We speculate that this phenomenon might be most
easily measured in cold atom experiments.

\section{Entropy principle in hydrodynamics \label{sub:Entropy-principle
copy(1)}}

\subsection{Single flavor case}

Let us start from the hydrodynamical system with only one flavor of quark of
electric charge $Q$. The energy-momentum conservation and current
conservation yield 
\begin{align}
\partial _{\mu }T^{\mu \nu }& =QF^{\nu \lambda }j_{\lambda },  \notag \\
\partial _{\mu }j^{\mu }& =0,  \label{eq:conservation_01}
\end{align}%
where $T^{\mu \nu }$ is the energy-momentum tensor, $j^{\mu }$ is the quark
current and $F^{\nu \lambda }$ is the electromagnetic field strength tensor.
The long wave length physics can be systematically described by the
expansion of space-time derivatives%
\begin{align}
T^{\mu \nu }& =T_{(0)}^{\mu \nu }+\varepsilon T_{(1)}^{\mu \nu }+\varepsilon
^{2}T_{(2)}^{\mu \nu }+\cdots ,  \notag \\
j^{\mu }& =j_{(0)}^{\mu }+\varepsilon j_{(1)}^{\mu }+\varepsilon
^{2}j_{(2)}^{\mu }+\cdots ,
\end{align}%
where we have used the parameter $\varepsilon $ to keep track of the
expansion and we will set $\varepsilon =1$ at the end. $F^{\nu \lambda }$ is
counted as $\mathcal{O}(\varepsilon )$. We will then assume the system is
isotropic and homogeneous in thermal equilibrium so there is no special
directions or intrinsic length scales macroscopically. We also assume the
underlying microscopic theory satisfies parity, charge conjugation and time
reversal symmetries such that the antisymmetric tensor $\varepsilon ^{\mu
\nu \alpha \beta }$ does not contribute to $T^{\mu \nu }$ and $j^{\mu }$.
Also, we assume the system is fluid-like, describable by one (and only one)
velocity field (the conserved charged is assumed to be not broken
spontaneously, otherwise the superfluid velocity needs to be introduced as
well). Also, at $\mathcal{O}(\varepsilon ^{0})$, the system is in local
thermal equilibrium, i.e. the system is in equilibrium in the comoving frame
where the fluid velocity is zero. With these assumptions, we can parametrize 
\begin{align}
T_{(0)}^{\mu \nu }& =\left( \epsilon +P\right) u^{\mu }u^{\nu }-Pg^{\mu \nu
},  \notag \\
j_{(0)}^{\mu }& =nu^{\mu },  \notag \\
T_{(1)}^{\mu \nu }& =\Pi \left( u^{\mu }u^{\nu }-g^{\mu \nu }\right) +\pi
^{\mu \nu }+h^{\mu }u^{\nu }+h^{\nu }u^{\mu },  \notag \\
j_{(1)}^{\mu }& =\nu ^{\mu },  \label{T}
\end{align}%
where $g^{\mu \nu }=$diag($+,-,-,-$ ) and $\epsilon $, $P$ and $n$ are the
energy density, pressure and number density, respectively. The fluid
velocity $u^{\mu }=(u^{0},\boldsymbol{u})=(1,\boldsymbol{v})/\sqrt{1-%
\boldsymbol{v}^{2}}$ and $u^{\mu }u_{\mu }=1.$ $\Pi $, $\pi ^{\mu \nu }$, $%
h^{\mu }$ and $\nu ^{\mu }$ are the bulk viscous pressure, shear viscous
tensor, heat flow vector and diffusion current. They satisfy the orthogonal
relations, $\pi ^{\mu \nu }u_{\nu }=\nu ^{\mu }u_{\mu }=h^{\mu }u_{\mu }=0$.

The covariant entropy flow is given by \cite{Israel:1979wp,Pu:2011vr} 
\begin{equation}
S^{\mu }=\beta Pu^{\mu }+\beta T^{\mu \nu }u_{\nu }-\beta \mu j^{\mu
}=su^{\mu }+\beta h^{\mu }-\beta \mu \nu ^{\mu },  \label{eq:entropy_flow_01}
\end{equation}%
where $\beta =1/T$ and $s=\beta (\epsilon +P-\mu n)$ is the entropy density.
Taking the space-time derivative of $S^{\mu }$, then using the Gibbs-Duhem
relation $d\epsilon =Tds+\mu dn$ and the conservation equations (\ref%
{eq:conservation_01}), we obtain the equation for entropy production: 
\begin{align}
\partial _{\mu }S^{\mu }& =-\nu ^{\mu }\left[ \partial _{\mu }(\beta \mu
)+\beta QE_{\mu }\right] +h^{\mu }\left( \partial _{\mu }\beta +\beta u^{\nu
}\partial _{\nu }u_{\mu }\right)  \notag \\
& +\beta \pi ^{\mu \nu }\partial _{\langle\mu }u_{\nu \rangle}-\beta \Pi
\partial \cdot u,  \label{div-s-1}
\end{align}%
where the symmetric traceless tensor $\partial _{\langle \mu }u_{\nu
\rangle} $ is defined by, 
\begin{equation}
\partial _{\langle\mu }u_{\nu \rangle}=\frac{1}{2}\left[ \Delta _{\mu \alpha
}\Delta _{\nu \beta }+\Delta _{\nu \alpha }\Delta _{\mu \beta }-\frac{2}{3}%
\Delta _{\mu \nu }\Delta _{\alpha \beta }\right] \partial ^{\alpha }u^{\beta
},
\end{equation}%
and where$\;\Delta ^{\mu \nu }=g^{\mu \nu }-u^{\mu }u^{\nu }$ and $E^{\mu
}\equiv F^{\mu \nu }u_{\nu }$ is the electric field in the comoving frame.

At $\mathcal{O}(\varepsilon )$, $\partial _{\mu }T^{\mu \nu }\simeq \partial
_{\mu }T_{(0)}^{\mu \nu }=0$. This equation yields 
\begin{equation}
\partial _{\mu }\beta +\beta u^{\nu }\partial _{\nu }u_{\mu }=\frac{n}{%
\epsilon +P}[\partial _{\mu }(\beta \mu )+\beta QE_{\mu }],
\label{eq:relation_01}
\end{equation}%
where we have used the thermodynamic equation $dP=\beta (\epsilon
+P)dT+nTd(\beta \mu )$. This identity simplifies Eq. (\ref{div-s-1}) to 
\begin{align}
\partial _{\mu }S^{\mu }& =-\left( \nu ^{\mu }-\frac{n}{\epsilon +P}h^{\mu
}\right) [\partial _{\mu }(\beta \mu )+\beta QE_{\mu }]  \notag \\
& +\beta \pi ^{\mu \nu }\partial _{\left\langle \mu \right. }u_{\left. \nu
\right\rangle }-\beta \Pi \partial \cdot u.  \label{eq:h_nu_02-1}
\end{align}

The second law of thermodynamics requires $\partial _{\mu }S^{\mu }\geq 0$.
It can be satisfied if, up to terms orthogonal to $\partial \cdot u$, $%
\partial _{\left\langle \mu \right. }u_{\left. \nu \right\rangle }$ and $%
[\partial _{\mu }(\beta \mu )+\beta QE_{\mu }]$, $\Pi $, $\pi ^{\mu \nu }$, $%
h^{\mu }$ and $\nu ^{\mu }$ have the following forms at $\mathcal{O}%
(\varepsilon )$: 
\begin{align}
\pi ^{\mu \nu }& =2\eta \partial ^{\left\langle \mu \right. }u^{\left. \nu
\right\rangle },  \notag \\
\Pi & =-\zeta \partial \cdot u,  \notag \\
\nu ^{\mu }-\frac{n}{\epsilon +P}h^{\mu }& =\lambda \Delta ^{\mu \nu }\left[
\partial _{\nu }(\beta \mu )+\beta QE_{\nu }\right] ,
\label{eq:definition_01}
\end{align}%
where $\Delta ^{\mu \nu }$ is inserted because $\nu ^{\mu }u_{\mu }=h^{\mu
}u_{\mu }=0$. The coefficients $\eta $, $\zeta $ and $\lambda $ are
transport coefficients with names of shear viscosity, bulk viscosity and
conductivity, respectively. The second law of thermodynamics requires these
transport coefficients to be non-negative.

On the right hand side of Eq. (\ref{eq:definition_01}), the three vectors $%
\partial _{\nu }\mu $, $\partial _{\nu }\beta $ and $E_{\nu }$ form a unique
combination and share the same transport coefficient $\lambda $ \cite%
{Israel:1979wp}. It is obtained by assuming $\partial _{\mu }T_{(0)}^{\mu
\nu }=0$ and $T_{(0)}^{\mu \nu }$ has the ideal fluid form described in Eq.(%
\ref{T}). In general, we do not expect this to be true in all systems (e.g.
a solid might not have the ideal fluid description) and hence there could be
more transport coefficients. Conventionally, the transport coefficients
corresponding to $\partial _{\nu }\mu $, $\partial _{\nu }\beta $ and $%
E_{\nu }$ are called particle diffusion, heat conductivity, and electric
conductivity, respectively.

In hydrodynamics, the choice of the velocity field is not unique. One could
choose $\mathbf{u}$ to align with the momentum density $T^{0i}\mathbf{\hat{i}%
}$ or the current $\mathbf{j}$, or their combinations. However, the system
should be invariant under the transformation $u_{\mu }\rightarrow u_{\mu
}^{^{\prime }}=u_{\mu }+\varepsilon \delta u_{\mu }$ as long as $u_{\mu
}^{^{\prime }2}=1$ is maintained (or $u^{\mu }\delta u_{\mu }=0$ at $%
\mathcal{O}(\varepsilon )$). Under this transformation, $h_{\mu }\rightarrow 
$ $h_{\mu }^{\prime }=h_{\mu }+\left( \epsilon +P\right) \delta u_{\mu }$
and $\nu _{\mu }\rightarrow $ $\nu _{\mu }^{\prime }=\nu _{\mu }+n\delta
u_{\mu }$. However, the entropy production equation (\ref{eq:h_nu_02-1})
remains invariant under this transformation.

In this paper, we will be working at the Landau frame with $\mathbf{u}$
proportional to the momentum density $T^{0i}\mathbf{\hat{i}}$ such that $%
T^{0i}=0$\ in the comoving frame. Then

\begin{equation}
\mathbf{h}=0,\quad \bm{\nu }=\lambda \lbrack -\bm{\nabla }(\beta \mu )+\beta
Q\mathbf{E}]  \label{aa}
\end{equation}%
from Eq. (\ref{eq:definition_01}). $\lambda $ is positive, the sign makes
sense for particle diffusion and electric conduction because the diffusion
is from high to low density and positively charged particles move along the $%
\mathbf{E}$ direction. However, heat conduction induces a flow from low to
high temperature! This result is counter intuitive. This is because $%
\bm{\nabla
}T$ induces a momentum flow $\mathbf{h}$. If we choose to boost the system
to the Landau frame where $\mathbf{h}=0$, then the physics is less
transparent. For particle diffusion and electric conduction this is not a
problem, because one could have particles and antiparticles moving in
opposite directions and still keep the net momentum flow zero.

The physics of heat conduction becomes clear in the Eckart frame where $%
\mathbf{u}$ is proportional to the current $\mathbf{j}$ and we have

\begin{equation}
\bm{\nu }=0,\quad \mathbf{h}=-\frac{\epsilon +P}{n}\lambda \lbrack -%
\bm{\nabla }(\beta \mu )+\beta Q\mathbf{E}].
\end{equation}%
In this frame, the direction of heat conduction is correct (while the
physics of particle diffusion and electric conduction become less
transparent). As expected, $\mathbf{h}$ stays finite when $\mu =Q=0$ but $%
\bm{\nabla }\beta \neq 0$.

\subsection{Multi-flavor case}

When the flavor of massless quarks is increased to $N_{f}$, then there are $%
N_{f}$ conserved currents $j_{a}^{\mu }$ (the conserved electric current is
just a combination of them). The hydrodynamical equations becomes 
\begin{align}
\partial _{\mu }T^{\mu \nu }& =\sum_{a=1}^{N_{f}}Q_{a}F^{\nu \lambda
}j_{a,\lambda },  \notag \\
\partial _{\mu }j_{a}^{\mu }& =0,\ a=1,2,\cdots ,N_{f}.
\label{eq:conservation_01-1}
\end{align}%
Then the entropy production yields 
\begin{eqnarray}
\partial _{\mu }S^{\mu } &=&-\sum_{a=1}^{N_{f}}\left( \nu _{a}^{\mu }-\frac{%
n_{a}}{\epsilon +P}h^{\mu }\right) \left[ \partial _{\mu }(\beta \mu
_{a})+\beta Q_{a}E_{\mu }\right] +\beta \pi ^{\mu \nu }\partial
_{\left\langle \mu \right. }u_{\left. \nu \right\rangle }-\beta \Pi \partial
\cdot u  \notag \\
&\geq &0
\end{eqnarray}%
Working in the Landau frame, we have%
\begin{equation}
\bm{\nu }_{a}=\sum_{b=1}^{N_{f}}\lambda _{ab}[-\bm{\nabla }(\beta \mu
_{b})+\beta Q_{b}\mathbf{E}].  \label{d1}
\end{equation}%
Our task is to compute the $\lambda $ matrix which can be achieved by
setting $\bm{\nabla }(\beta \mu _{b})\neq 0$ but $\mathbf{E}=0$. The second
law of thermodynamics dictates $\lambda $ being a positive definite matrix.

\section{Effective kinetic theory \label{sec:Effective-kinetic-theory}}

We will use the Boltzmann equation to compute our LL result of $\lambda $.
It has been shown that Boltzmann equation gives the same leading order
result as the Kubo formula in the coupling constant expansion in a weakly
coupled $\phi ^{4}$ theory \cite{Jeon:1994if,Hidaka:2010gh} and in hot QED 
\cite{Gagnon:2007qt}, provided the leading $T$ and $\mu $ dependence in
particle masses and scattering amplitudes are included. This conclusion is
expected to hold in perturbative QCD as well \cite{Arnold:2002zm}.

The Boltzmann equation of a quark gluon plasma describes the evolution of
the color and spin averaged distribution function $\tilde{f}_{p}^{i}(x)$ for
particle $i$ ($i=g,q_{a},\bar{q}_{a}$\ with $a=1.2...N_{f}$\ for gluon, $%
N_{f}$\ quarks and $N_{f}$\ anti-quarks): 
\begin{equation}
\frac{d\tilde{f}_{p}^{i}(x)}{dt}=\mathcal{\tilde{C}}_{i},  \label{eq:BE_01}
\end{equation}%
where $\tilde{f}_{p}^{i}(x)$ is a function of space-time $x^{\mu }=(t,%
\mathbf{x})$ and momentum $p^{\mu }=(E_{p},\mathbf{p})$.

For the LL calculation, we only need to consider two-particle scattering
processes denoted as $c_{1}c_{2}\rightarrow c_{3}c_{4}$. The collision term
has the form 
\begin{equation}
C_{c_{1}c_{2}\rightarrow c_{3}c_{4}}\equiv \int_{k_{1}k_{2}k_{3}}d\Gamma
_{c_{1}c_{2}\rightarrow c_{3}c_{4}}\left[ \tilde{f}_{k_{1}}^{c_{1}}\tilde{f}%
_{k_{2}}^{c_{2}}\tilde{F}_{p}^{c_{3}}\tilde{F}_{k_{3}}^{c_{4}}-\tilde{F}%
_{k_{1}}^{c_{1}}\tilde{F}_{k_{2}}^{c_{2}}\tilde{f}_{p}^{c_{3}}\tilde{f}%
_{k_{3}}^{c_{4}}\right] .  \label{definition of C ab-cd}
\end{equation}%
where $\tilde{F}^{g}=1+\tilde{f}^{g}$ and $\tilde{F}^{q(\bar{q})}=1-\tilde{f}%
^{q(\bar{q})}$ and 
\begin{equation}
d\Gamma _{c_{1}c_{2}\rightarrow c_{3}c_{4}}=\frac{1}{2E_{p}}%
|M_{c_{1}c_{2}\rightarrow c_{3}c_{4}}|^{2}\prod\limits_{i=1}^{3}\frac{%
d^{3}k_{i}}{(2\pi )^{3}2E_{k_{i}}}(2\pi )^{4}\delta
^{(4)}(k_{1}+k_{2}-k_{3}-p),  \label{gamma ab-cd}
\end{equation}%
where $|M_{c_{1}c_{2}\rightarrow c_{3}c_{4}}|^{2}$ is the matrix element
squared with all colors and helicities of the initial and final states
summed over. The scattering amplitudes can be regularized by hard thermal
loop propagators and in this paper we use the same scattering amplitudes as
in Ref. \cite{Arnold:2003zc} (see also Table I of Ref. \cite{Chen:2012jc}).
Then the collision term for a quark of flavor $a$ is 
\begin{align}
N_{q}\mathcal{\tilde{C}}_{q_{a}}& =\frac{1}{2}C_{q_{a}q_{a}\leftrightarrow
q_{a}q_{a}}+C_{q_{a}\bar{q}_{a}\leftrightarrow q_{a}\bar{q}_{a}}+\frac{1}{2}%
C_{gg\leftrightarrow q_{a}\bar{q}_{a}}+C_{q_{a}g\leftrightarrow q_{a}g} 
\notag \\
& +\sum\limits_{b,b\neq a}^{N_{f}-1}(C_{q_{a}q_{b}\leftrightarrow
q_{a}q_{b}}+C_{q_{a}\bar{q}_{b}\leftrightarrow q_{a}\bar{q}_{b}}+C_{q_{b}%
\bar{q}_{b}\leftrightarrow q_{a}\bar{q}_{a}}),
\end{align}%
where $N_{q}=2\times 3=6$ is the quark helicity and color degeneracy factor
and the factor $1/2$ is included when the initial state is formed by two
identical particles. Similarly, 
\begin{equation}
N_{g}\mathcal{\tilde{C}}_{g}=\frac{1}{2}C_{gg\leftrightarrow
gg}+\sum\limits_{a=1}^{N_{f}}(C_{gq_{a}\leftrightarrow gq_{a}}+C_{g\bar{q}%
_{a}\leftrightarrow g\bar{q}_{a}}+C_{q_{a}\bar{q}_{a}\leftrightarrow gg}),
\label{C22g}
\end{equation}%
where $N_{g}=2\times 8=16$ is the gluon helicity and color degeneracy
factor. In equilibrium, the distributions are denoted as $f^{q_{a}(\bar{q}%
_{a})}$ and $f^{g}$, with 
\begin{align}
f_{p}^{g}& =\frac{1}{e^{u\cdot p/T}-1},  \label{distribution g} \\
f_{p}^{q_{a}(\bar{q}_{a})}& =\frac{1}{e^{(u\cdot p\mp \mu _{a})/T}+1},
\label{distribution q}
\end{align}%
where $T$ is the temperature, $u$ is the fluid four velocity and $\mu _{a}$
is the chemical potential for the quark of flavor $a$. They are all space
time dependent.

The thermal masses of gluon and quark/anti-quark for external states (the
asymptotic masses) can be computed via \cite%
{Arnold:2002zm,Mrowczynski:2000ed} 
\begin{align}
m_{g}^{2}& =\sum_{i}N_{i}C_{i}\frac{2g^{2}}{d_{A}}\int \frac{d^{3}p}{(2\pi
)^{3}2E_{p}}f_{p}^{i},  \label{mg} \\
m_{q}^{2}& =m_{\bar{q}}^{2}=2C_{F}g^{2}\int \frac{d^{3}p}{(2\pi )^{3}2E_{p}}%
(2f_{p}^{g}+f_{p}^{q}+f_{p}^{\bar{q}}),  \label{thermal mass g}
\end{align}%
where $d_{A}=8$, $C_{g}=C_{A}=3$, and $C_{q(\bar{q})}=C_{F}=4/3$. This
yields 
\begin{align}
m_{g}^{2}& =\frac{C_{A}}{6}g^{2}T^{2}+\sum_{a=1}^{N_{f}}\frac{C_{F}}{16}%
g^{2}(T^{2}+\frac{3}{\pi ^{2}}\mu _{a}^{2}), \\
m_{q_{a}}^{2}& =\frac{1}{4}C_{F}g^{2}\left( T^{2}+\frac{\mu _{a}^{2}}{\pi
^{2}}\right) ,  \label{eqi thermal mass g}
\end{align}%
where we have set $E_{p}=|\mathbf{p}|$ in the integrals on the right hand
sides of Eqs.\ (\ref{mg}) and (\ref{thermal mass g}). The difference from
non-vanishing masses is of higher order. In this work, we only need the fact
that the thermal masses are proportional to $g^{2}$ for the LL results.

\subsection{Linearized Boltzmann equation}

Matching to the derivative expansion in hydrodynamics, we expand the
distribution function of particle $i$ as a local equilibrium distribution
plus a correction 
\begin{equation}
\tilde{f}_{p}^{i}(x)=f_{p}^{i}-\varepsilon f_{p}^{i}(1\mp f_{p}^{i})\chi
^{i},  \label{eq:expansion_01}
\end{equation}%
where the upper/lower sign corresponds to the femion/boson distribution.
Inserting Eq. (\ref{eq:expansion_01}) into Eq. (\ref{eq:BE_01}), we can
solve the linearized Boltzmann equation by keeping linear terms in
space-time derivatives. Here we neglect the viscous terms related to $%
\partial _{\mu }u_{\nu }$ in $\chi ^{i}$\ and consider only the $\bm{\nabla }%
(\beta \mu _{a})$ terms.

At the zeroth order, $\mathcal{O}(\varepsilon ^{0})$, the system is in local
thermal equilibrium and the Boltzmann equation (\ref{eq:BE_01}) is
satisfied, $\mathcal{\tilde{C}}[f_{p}^{i}]=0$. At $\mathcal{O}(\varepsilon )$%
, the left hand side of the Boltzmann equation yields 
\begin{equation}
\frac{df_{p}^{g}}{dt}=-\beta f_{p}^{g}F_{p}^{g}\sum_{a=1}^{N_{f}}\left[ 
\frac{n_{a}T}{\epsilon +P}\mathbf{p}\cdot \bm{\nabla }(\beta \mu _{a})\right]
,  \label{b1}
\end{equation}%
and 
\begin{equation}
\frac{df_{p}^{q_{a}(\bar{q}_{a})}}{dt}=-\beta f_{p}^{q_{a}(\bar{q}%
_{a})}F_{p}^{q_{a}(\bar{q}_{a})}\sum_{b=1}^{N_{f}}\left( \frac{n_{b}T}{%
\epsilon +P}\mp \frac{T}{E_{p}^{q_{a}(\bar{q}_{a})}}\delta _{ab}\right) 
\mathbf{p}\cdot \bm{\nabla }(\beta \mu _{b}).  \label{b2}
\end{equation}%
To derive this result, we have used $\partial _{\mu }u^{0}=0$ in the local
fluid rest frame where $u^{\mu }=(1,0,0,0)$ and $\partial _{\mu
}T_{(0)}^{\mu \nu }=0$ and $\partial _{\mu }j_{a(0)}^{\mu }=0$ which yields%
\begin{align}
\frac{\partial \epsilon }{\partial t}& =-\left( \epsilon +p\right) %
\bm{\nabla }\cdot \mathbf{u}  \label{energy density time derivation} \\
\frac{\partial \mathbf{u}}{\partial t}& =-\frac{\bm{\nabla }P}{\epsilon +p} 
\notag
\end{align}%
and%
\begin{equation}
\frac{\partial n_{a}}{\partial t}=-n_{a}\bm{\nabla }\cdot \mathbf{u}.
\label{quark number density time derivation}
\end{equation}%
And then by applying thermodynamic relations, we can replace the time
derivatives of $T$ , $\mu $ and $\mathbf{u}$ with spatial derivatives: 
\begin{align}
\frac{\partial T}{\partial t}& =-T\left( \frac{\partial P}{\partial \epsilon 
}\right) _{n}\bm{\nabla }\mathbf{\cdot u},  \notag \\
\frac{\partial \mu }{\partial t}& =-\left[ \mu \left( \frac{\partial P}{%
\partial \epsilon }\right) _{n}+\left( \frac{\partial P}{\partial n}\right)
_{\epsilon }\right] \bm{\nabla }\mathbf{\cdot u},  \label{dT/dt} \\
\frac{\partial \mathbf{u}}{\partial t}& =-\beta \bm{\nabla }%
T-\sum_{a=1}^{N_{f}}\frac{n_{a}T}{\epsilon +p}\bm{\nabla }\left( \frac{\mu
_{a}}{T}\right) .  \notag
\end{align}%
Those relations lead to Eqs.(\ref{b1},\ref{b2}).

To get the right hand side of the Boltzmann equation at $\mathcal{O}%
(\varepsilon )$, we parametrize $\chi ^{i}$ of Eq. (\ref{eq:expansion_01})
as 
\begin{align}
\chi ^{i}& =\beta \sum_{a=1}^{N_{f}}\mathbf{A}^{ia}(p)\cdot \bm{\nabla }%
(\beta \mu _{a}),  \notag \\
\mathbf{A}^{ia}(p)& =A^{ia}(|\mathbf{p}|)\hat{\mathbf{p}}.  \label{c1}
\end{align}%
The matrix $A^{ia}$ is $\left( 2N_{f}+1\right) \times N_{f}$. We will see
there are $\left( 2N_{f}+1\right) \times N_{f}$ equations to constrain them.

For each Boltzmann equation, we have a linear combination of $N_{f}$ terms
of $\bm{\nabla }(\beta \mu _{a})$. Since each $\bm{\nabla }(\beta \mu _{a})$
is linearly independent to each other, thus there are $N_{f}$ equations for
each Boltzmann equation. Totally we have $2N_{f}+1$ Boltzmann equations,
thus we have $\left( 2N_{f}+1\right) \times N_{f}$ equations to solve for $%
A^{ia}$. These equations are 
\begin{equation}
\frac{n_{a}\mathbf{p}}{\epsilon +P}=\beta \frac{1}{f_{p}^{g}F_{p}^{g}}\frac{1%
}{N_{g}}\left[ \frac{1}{2}\mathbf{C}_{gg\text{ }\rightarrow
gg}^{a}+\sum\limits_{c=1}^{N_{f}}\left( \mathbf{C}_{gq_{c}\text{ }%
\rightarrow gq_{c}}^{a}+\mathbf{C}_{g\bar{q}_{c}\rightarrow g\bar{q}%
_{c}}^{a}+\mathbf{C}_{q_{c}\bar{q}_{c}\rightarrow gg}^{a}\right) \right] ,
\label{eq:constr-g}
\end{equation}%
\begin{align}
\left( \frac{n_{a}}{\epsilon +P}-\frac{1}{E_{p}^{q_{b}}}\delta _{ab}\right) 
\mathbf{p}& =\frac{\beta }{f_{p}^{q_{b}}F_{p}^{q_{b}}}\frac{1}{N_{q}}\left[ 
\frac{1}{2}\mathbf{C}_{q_{b}q_{b}\rightarrow q_{b}q_{b}}^{a}+\mathbf{C}%
_{q_{b}\bar{q}_{b}\rightarrow q_{b}\bar{q}_{b}}^{a}+\frac{1}{2}\mathbf{C}_{g%
\text{ }g\text{ }\rightarrow q_{b}\bar{q}_{b}}^{a}+\mathbf{C}%
_{q_{b}g\rightarrow q_{b}g}^{a}\right.  \notag \\
& +\left. \sum\limits_{c,c\neq b}^{N_{f}-1}\left( \mathbf{C}%
_{q_{b}q_{c}\rightarrow q_{b}q_{c}}^{a}+\mathbf{C}_{q_{b}\bar{q}%
_{c}\rightarrow q_{b}\bar{q}_{c}}^{a}+\mathbf{C}_{q_{c}\bar{q}%
_{c}\rightarrow q_{b}\bar{q}_{b}}^{a}\right) \right] ,  \label{eq:constr-q}
\end{align}%
and 
\begin{align}
\left( \frac{n_{a}}{\epsilon +P}+\frac{1}{E_{p}^{\bar{q}_{b}}}\delta
_{ab}\right) \mathbf{p}& =\frac{\beta }{f_{p}^{\bar{q}_{b}}F_{p}^{\bar{q}%
_{b}}}\frac{1}{N_{q}}\left[ \frac{1}{2}\mathbf{C}_{\bar{q}_{b}\bar{q}%
_{b}\rightarrow \bar{q}_{b}\bar{q}_{b}}^{a}+\mathbf{C}_{\bar{q}%
_{b}q_{b}\rightarrow \bar{q}_{b}q_{b}}^{a}+\frac{1}{2}\mathbf{C}_{gg\text{ }%
\rightarrow \bar{q}_{b}q_{b}}^{a}+\mathbf{C}_{\bar{q}_{b}g\rightarrow \bar{q}%
_{b}g}^{a}\right.  \notag \\
& +\left. \sum\limits_{c,c\neq b}^{N_{f}-1}\left( \mathbf{C}_{\bar{q}_{b}%
\bar{q}_{c}\rightarrow \bar{q}_{b}\bar{q}_{c}}^{a}+\mathbf{C}_{\bar{q}%
_{b}q_{c}\rightarrow \bar{q}_{b}q_{c}}^{a}+\mathbf{C}_{\bar{q}%
_{c}q_{c}\rightarrow \bar{q}_{b}q_{b}}^{a}\right) \right] ,
\label{eq:constr-qb}
\end{align}%
where 
\begin{align}
\mathbf{C}_{c_{1}c_{2}\rightarrow c_{3}c_{4}}^{a}(p)& \equiv
\int_{k_{1}k_{2}k_{3}}d\Gamma _{c_{1}c_{2}\rightarrow
c_{3}c_{4}}f^{c_{1}}f^{c_{2}}F^{c_{3}}F^{c_{4}}  \notag \\
& \times \left[ \mathbf{A}^{c_{1}a}(k_{1})+\mathbf{A}^{c_{2}a}(k_{2})-%
\mathbf{A}^{c_{3}a}(k_{3})-\mathbf{A}^{c_{4}a}(p)\right] .
\end{align}%
Formally we can rewrite these linearized Boltzmann equations in a compact
form, 
\begin{equation}
\left\vert \mathbf{S}^{a}\right\rangle =\mathcal{C}_{\lambda }\left\vert 
\mathbf{A}^{a}\right\rangle ,  \label{linearized equation}
\end{equation}%
where $\left\vert \mathbf{S}^{a}\right\rangle $ and $\left\vert \mathbf{A}%
^{a}\right\rangle $ are both vectors of $\left( 2N_{f}+1\right) $ components
and $\mathcal{C}_{\lambda }$ is a $\left( 2N_{f}+1\right) \times \left(
2N_{f}+1\right) $ matrix.

\subsection{Conductivity matrix}

In the kinetic theory, the quark current of flavor $a$ is 
\begin{equation}
j_{a}^{\mu }=N_{q}\int \frac{d^{3}p}{\left( 2\pi \right) ^{3}}\frac{p^{\mu }%
}{E^{q_{a}}}\left( \tilde{f}_{p}^{q_{a}}-\tilde{f}_{p}^{\bar{q}_{a}}\right) .
\end{equation}%
Expanding this expression to $\mathcal{O}(\varepsilon )$ and matching it to
Eq.(\ref{d1}), we have%
\begin{equation}
\lambda _{ab}=\frac{N_{q}\beta }{3}\int \frac{d^{3}p}{(2\pi )^{3}}\frac{1}{%
E^{q_{a}}}\left( f^{q_{a}}F^{q_{a}}\mathbf{p\cdot }\mathbf{A}^{q_{a}b}-f^{%
\bar{q}_{a}}F^{\bar{q}_{a}}\mathbf{p\cdot }\mathbf{A}^{\bar{q}_{a}b}\right) .
\label{e1}
\end{equation}%
Since we are working in the Landau frame, we should impose the
Landau-Lifshitz condition 
\begin{equation}
0=T^{0j}=-\sum_{i}N_{i}\int \frac{d^{3}p}{\left( 2\pi \right) ^{3}}%
f^{i}F^{i}\chi ^{i}p^{j}.
\end{equation}%
This implies 
\begin{equation}
\sum_{i}N_{i}\int \frac{d^{3}p}{\left( 2\pi \right) ^{3}}f^{i}F^{i}\mathbf{%
p\cdot A}^{ia}=0.
\end{equation}%
We can use these constraints to rewrite Eq.(\ref{e1}) as%
\begin{align}
\lambda _{ab}& =-\frac{\beta }{3}\left\{ N_{g}\int \frac{d^{3}p}{(2\pi )^{3}}%
f^{g}F^{g}\frac{n_{a}}{\epsilon +P}\mathbf{p\cdot A}^{gb}\right.  \notag \\
& +\sum_{c=1}^{N_{f}}N_{q}\int \frac{d^{3}p}{(2\pi )^{3}}f^{q_{c}}F^{q_{c}}%
\left( \frac{n_{a}}{\epsilon +P}-\frac{1}{E^{q_{a}}}\delta _{ca}\right) 
\mathbf{p}\cdot \mathbf{A}^{q_{c}b}  \notag \\
& \left. +\sum_{c=1}^{N_{f}}N_{q}\int \frac{d^{3}p}{(2\pi )^{3}}f^{\bar{q}%
_{c}}F^{\bar{q}_{c}}\left( \frac{n_{a}}{\epsilon +P}+\frac{1}{E^{q_{a}}}%
\delta _{ca}\right) \mathbf{p}\cdot \mathbf{A}{}^{\bar{q}_{c}b}\right\} .
\label{01}
\end{align}%
This form can be schematically written as 
\begin{equation}
\lambda _{ab}=\langle \mathbf{A}^{b}|\mathbf{S}^{a}\rangle =\langle \mathbf{A%
}^{b}|\mathcal{C}_{\lambda }\left\vert \mathbf{A}^{a}\right\rangle ,
\label{f3}
\end{equation}%
where we have used Eq.(\ref{linearized equation}) for the second equality.
More explicitly, 
\begin{align}
\lambda _{ab}& =\frac{\beta ^{2}}{24}\left( D_{gg\rightarrow
gg}^{ab}+\sum_{c=1}^{N_{f}}D_{q_{c}q_{c}\rightarrow
q_{c}q_{c}}^{ab}+\sum_{c=1}^{N_{f}}D_{\bar{q}_{c}\bar{q}_{c}\rightarrow \bar{%
q}_{c}\bar{q}_{c}}^{ab}\right)  \notag \\
& +\frac{\beta ^{2}}{6}\sum_{c=1}^{N_{f}}\left( D_{q_{c}\bar{q}%
_{c}\rightarrow gg}^{ab}+D_{gq_{c}\rightarrow gq_{c}}^{ab}+D_{g\bar{q}%
_{c}\rightarrow g\bar{q}_{c}}^{ab}+D_{q_{c}\bar{q}_{c}\rightarrow q_{c}\bar{q%
}_{c}}^{ab}\right)  \notag \\
& +\frac{\ \beta ^{2}}{12}\sum_{\substack{ c,d=1  \\ c\neq d}}^{N_{f}}\left(
D_{q_{c}q_{d}\rightarrow q_{c}q_{d}}^{ab}+D_{\bar{q}_{c}\bar{q}%
_{d}\rightarrow \bar{q}_{c}\bar{q}_{d}}^{ab}+2D_{q_{c}\bar{q}_{d}\rightarrow
q_{c}\bar{q}_{d}}^{ab}+2D_{q_{c}\bar{q}_{c}\rightarrow q_{d}\bar{q}%
_{d}}^{ab}\right) ,  \label{f1}
\end{align}%
where 
\begin{align}
D_{c_{1}c_{2}\rightarrow c_{3}c_{4}}^{ab}\equiv & \int \prod_{i=1}^{4}\frac{%
d^{3}p_{i}}{(2\pi )^{3}2E_{i}}(2\pi )^{4}\delta ^{4}(k_{1}+k_{2}-k_{3}-k_{4})
\notag \\
& \times |M_{c_{1}c_{2}\rightarrow
c_{3}c_{4}}|^{2}f_{k_{1}}^{c_{1}}f_{k_{2}}^{c_{2}}F_{k_{3}}^{c_{3}}F_{k_{4}}^{c_{4}}
\notag \\
& \times \left[ \mathbf{A}^{c_{1}a}(k_{1})+\mathbf{A}^{c_{2}a}(k_{2})-%
\mathbf{A}^{c_{3}a}(k_{3})-\mathbf{A}^{c_{4}a}(k_{4})\right]  \notag \\
& \cdot \left[ \mathbf{A}^{c_{1}b}(k_{1})+\mathbf{A}^{c_{2}b}(k_{2})-\mathbf{%
A}^{c_{3}b}(k_{3})-\mathbf{A}^{c_{4}b}(k_{4})\right] .  \label{f2}
\end{align}

From Eq.(\ref{linearized equation}), it is clear that if 
\begin{equation}
\mathbf{A}_{0}^{ia}(p)=\mathbf{p},  \label{0}
\end{equation}%
then from momentum conservation this implies 
\begin{equation}
\mathcal{C}_{\lambda }\left\vert \mathbf{A}_{0}^{a}\right\rangle =0.
\end{equation}%
Those modes are called zero modes (denoted by the subscribe $0$ in Eq.(\ref%
{0})). They would have been a problem for Eq.(\ref{f3}) unless $\langle 
\mathbf{S}^{a}|\mathbf{A}_{0}^{a}\rangle =0$, but this is guaranteed from
the total momentum conservation at $\mathcal{O}(\varepsilon )$,%
\begin{equation}
\frac{d}{dt}\sum_{i}\int dp^{3}\mathbf{p}f_{p}^{i}(x)=0,
\end{equation}%
and Eqs.(\ref{b1},\ref{b2}). Thus, we can just solve for $\left\vert \mathbf{%
A}^{a}\right\rangle $ in Eq.(\ref{f3}) by discarding the zero modes.

From Eqs.(\ref{f1}) and (\ref{f2}), we can see easily that $\lambda
_{ab}=\lambda _{ba}$. This is a manifestation of the Onsager relation which
appears when particle scattering is symmetric under the time-reversal
transformation. We can also see that $\lambda $ is positive definite.

\section{The Leading-Log Results with Identical Chemical Potentials\label%
{sec:Numerical-results}}

Now we are ready to solve the conductivity matrix $\lambda $. Our strategy
to solve for $\lambda _{ab}$ is to make use of Eq.(\ref{f3}) to solve for $|%
\mathbf{A}^{a}\rangle $ from $\lambda _{aa}$ (no summation over $a$). Once
all the $|\mathbf{A}^{a}\rangle $ are obtained, $\lambda _{ab}$ can be
computed. Also, in solving for $\lambda _{aa}$, one can use the standard
algorithm to systematically approach the answer\ from below \cite%
{Chen:2011km}. The dependence on the strong coupling constant is similar to
that in shear viscosity---it is inversely proportional to the scattering
rate which scales as $g^{4}\ln g^{-1}$ with the $\ln g^{-1}$ dependence
coming from regularizing the collinear infrared singularity by the thermal
masses of quarks or gluons. $\lambda $ is of mass dimension two, thus we
will present our result in the normalized conductivity 
\begin{equation}
\tilde{\lambda}\equiv \lambda \beta ^{2}g^{4}\ln g^{-1}
\end{equation}%
such that $\tilde{\lambda}$ is dimensionless and coupling constant
independent.

For simplicity, we will concentrate on the linear response of a thermal
equilibrium system with all fermion chemical potentials to be identical,
i.e. $\mu _{a}=\mu $ for all $a$'s but each $\bm{\nabla }\left( \beta \mu
_{a}\right) $ could be varied independently. This symmetry makes all the
diagonal matrix elements (denoted as $\lambda _{qq}$) identical and all the
off-diagonal ones (denoted as $\lambda _{qq^{\prime }}$) identical. $\lambda
_{qq}$ and $\lambda _{qq^{\prime }}$ are even in $\mu $ (and so are $\tilde{%
\lambda}_{qq}$ and $\tilde{\lambda}_{qq^{\prime }}$) because our microscopic
interaction (in vacuum) is invariant under charge conjugation, thus $\lambda 
$ should be invariant under $\mu _{a}\rightarrow -\mu _{a}$.

It is easy to diagonalize $\lambda $. One eigenvalue is%
\begin{equation}
\lambda _{+}/N_{f}\equiv \lambda _{qq}+(N_{f}-1)\lambda _{qq^{\prime }},
\label{eq:lambda_+}
\end{equation}%
corresponding to the conductivity of the flavor singlet total quark current (%
$\lambda _{+}$ is the total quark current conductivity)%
\begin{equation}
\bm{\nu }=\sum_{a=1}^{N_{f}}\bm{\nu }_{a}=-\lambda _{+}\sum_{a=1}^{N_{f}}%
\frac{\bm{\nabla }(\beta \mu _{a})}{N_{f}}.
\end{equation}%
The other $(N_{f}-1)$ eigenvalues are degenerate with the value 
\begin{equation}
\lambda _{-}\equiv \lambda _{qq}-\lambda _{qq^{\prime }}.
\label{eq:lambda_-}
\end{equation}%
They are the conductivities of the flavor non-singlet currents

\begin{equation}
\Delta \bm{\nu }_{a}\equiv \bm{\nu }_{a}-\bm{\nu }_{1}=-\lambda _{-}%
\bm{\nabla }[\beta (\mu _{a}-\mu _{1})],
\end{equation}%
with $a=2,...,N_{f}$.

\begin{figure}[tbp]
\begin{centering}
\includegraphics[scale=0.7]{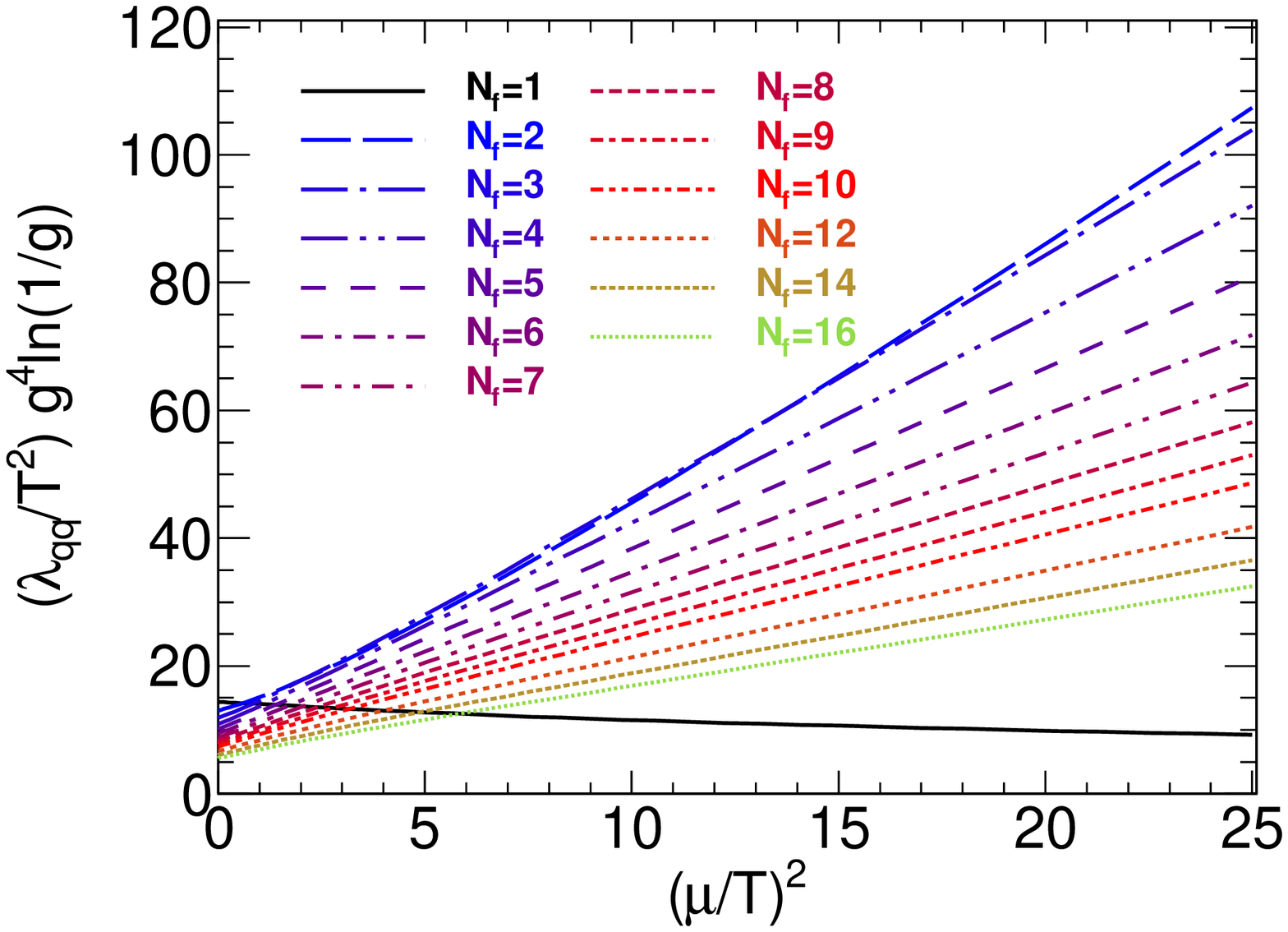}
\par\end{centering}
\par
\begin{centering}
\includegraphics[scale=0.7]{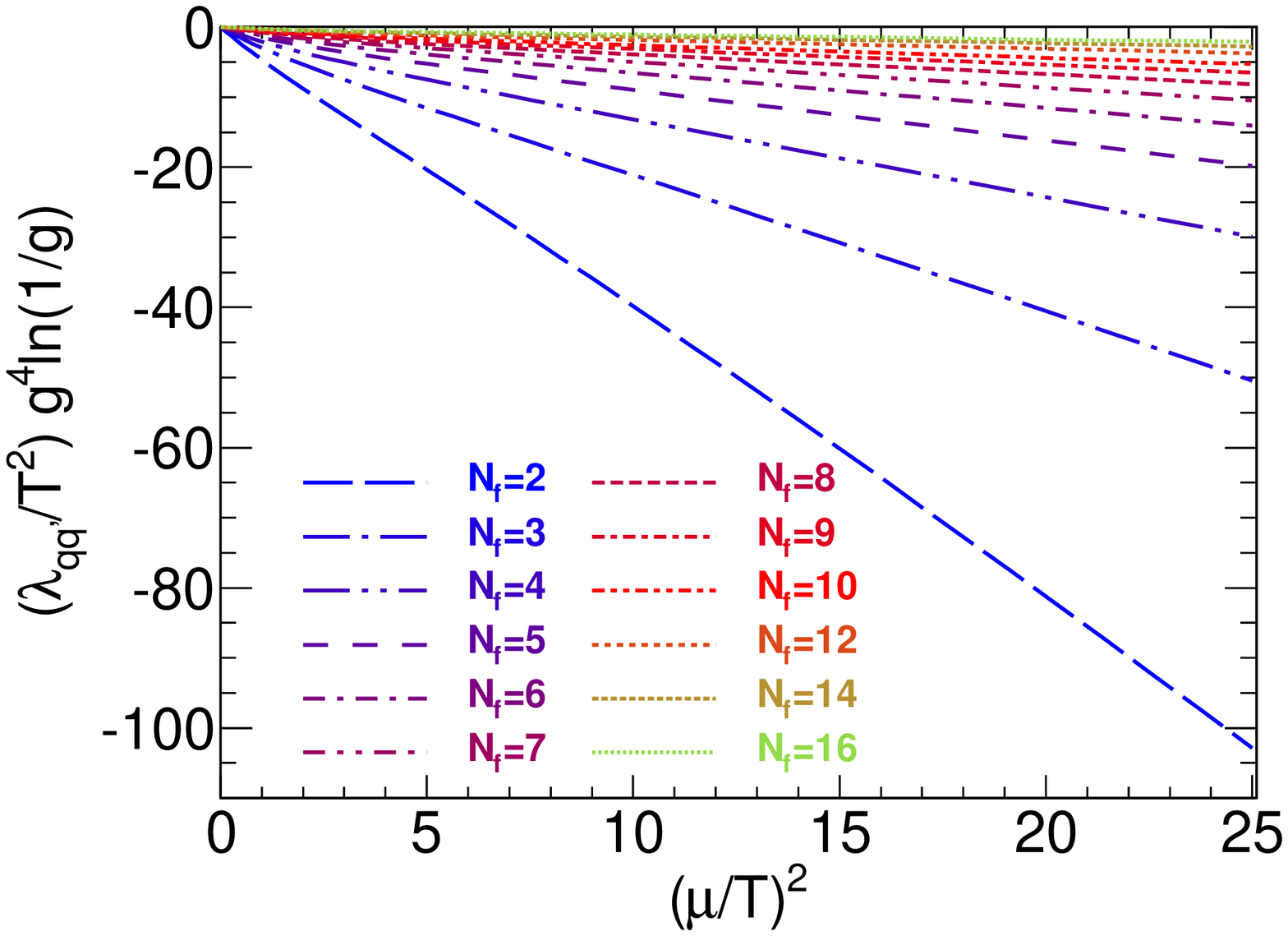}
\par\end{centering}
\caption{(color online). The normalized diagonal conductivity $\tilde{%
\protect\lambda}_{qq}$ (upper panel) and off-diagonal conductivity $\tilde{%
\protect\lambda}_{qq^{\prime }}$ (lower panel) as functions of $(\protect\mu %
/T)^{2}$ for different $N_{f}$. }
\label{fig. 1}
\end{figure}

\begin{figure}[tbp]
\begin{centering}
\includegraphics[scale=0.7]{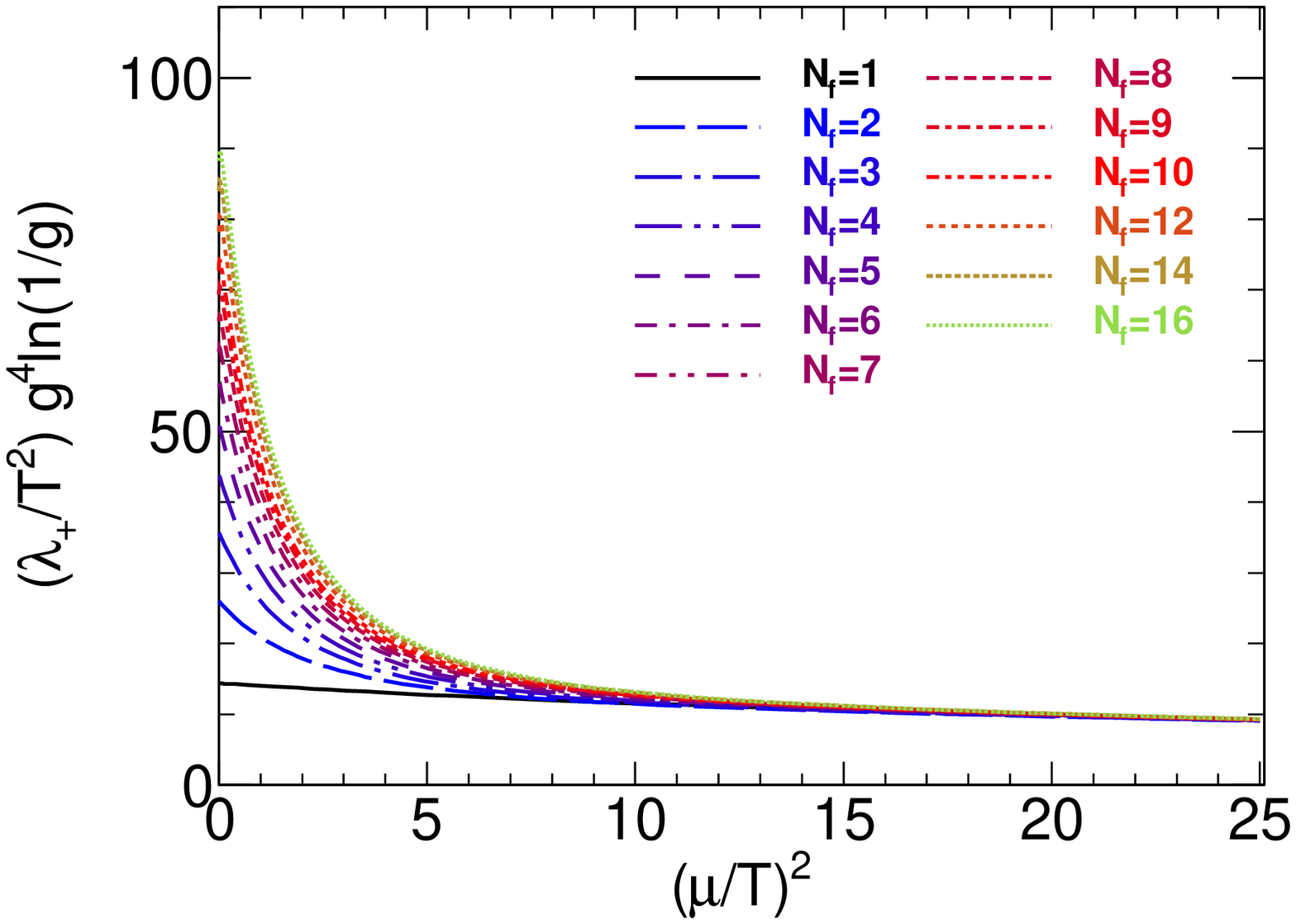}
\par\end{centering}
\par
\begin{centering}
\includegraphics[scale=0.7]{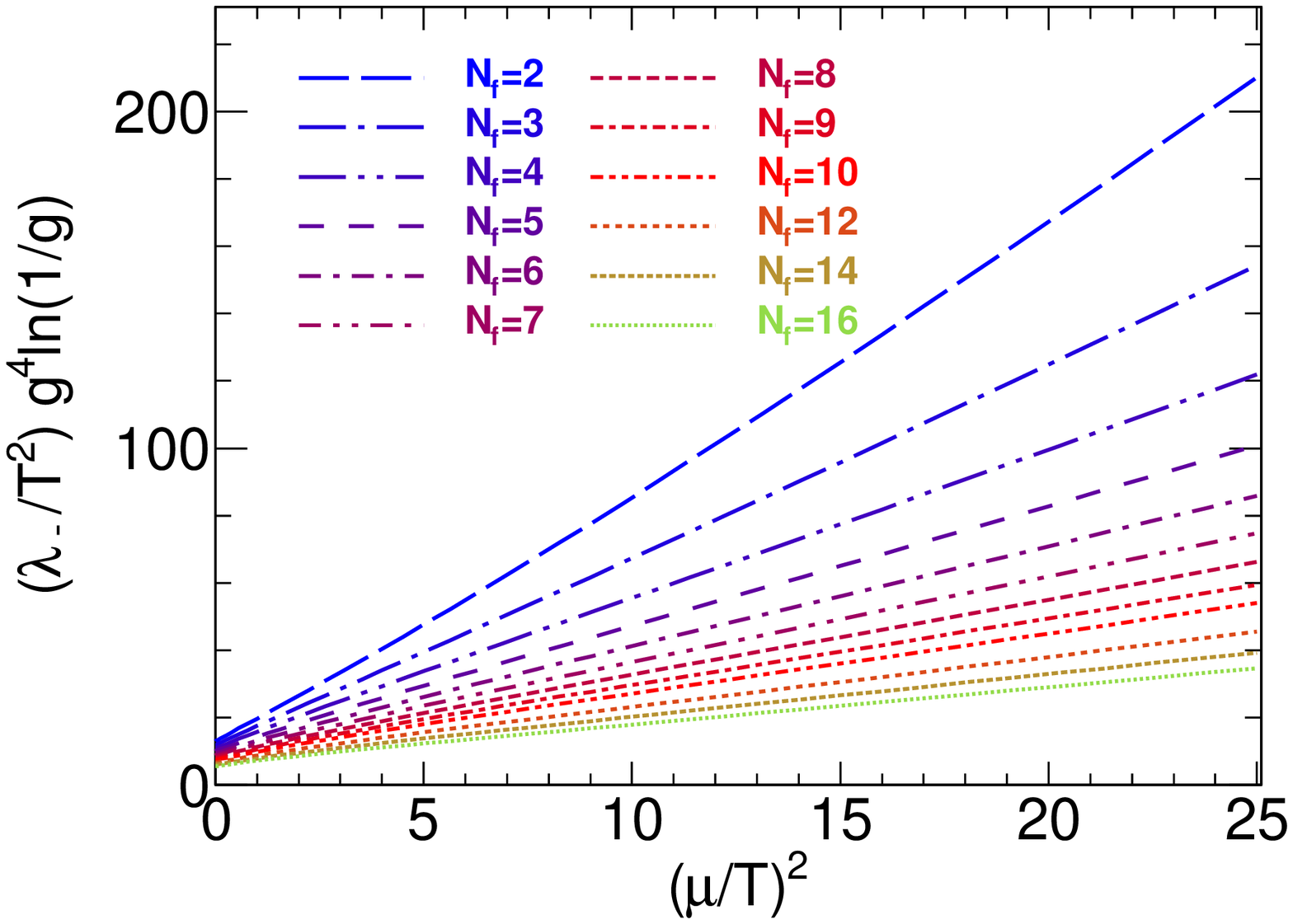}
\par\end{centering}
\par
\centering{}
\caption{(color online). The normalized conductivities $\tilde{\protect%
\lambda}_{+}$ (upper panel) and $\tilde{\protect\lambda}_{-}$ (lower panel)
as functions of $(\protect\mu /T)^{2}$ for different $N_{f}$ . }
\label{fig. 2}
\end{figure}

$\tilde{\lambda}_{qq}$ and $\tilde{\lambda}_{qq^{\prime }}$ are shown as
functions of $(\mu /T)^{2}$ in Fig. \ref{fig. 1} for various $N_{f}$ with $%
N_{f}\leq 16$ such that the system is asymptotically free, while $\tilde{%
\lambda}_{+}$ and $\tilde{\lambda}_{-}$ are shown in Fig. \ref{fig. 2} (note
that there is no $\tilde{\lambda}_{qq^{\prime }}$ or $\tilde{\lambda}_{-}$
for $N_{f}=1$). The fact that the matrix $\lambda $ is positive definite
makes $\tilde{\lambda}_{qq}$, $\tilde{\lambda}_{+}$ and $\tilde{\lambda}_{-}$
positive, but it imposes no constraint on the sign of $\tilde{\lambda}%
_{qq^{\prime }}$.

When $(\mu /T)^{2}\rightarrow 0$, we can expand $\tilde{\lambda}%
_{qq}=a_{0}+a_{1}(\mu /T)^{2}+...$, and $\tilde{\lambda}_{qq^{\prime
}}=a_{0}^{\prime }+a_{1}^{\prime }(\mu /T)^{2}+...$. We find $a_{0}^{\prime
}=0$ for all $N_{f}$ while the values of $a_{0},a_{1}$ and $a_{1}^{\prime }$
for different $N_{f}$ are tabulated in Table. \ref{power expansion
coefficients lambdaqq}. Our result for $a_{0}$ agrees within $0.1\%$ to that
of Arnold, Moore and Yaffe (AMY) calculated up to $N_{f}=6$ listed in Table
III of Ref. \cite{Arnold:2000dr}.

The $a_{0}^{\prime }=0$ property is due to a bigger symmetry enjoyed by the
LL results: if we just change all the quarks of flavor $a$ into anti-quarks
while the rest of the system stays the same, then as far as collision is
concerned, the other quarks and the gluons will not feel any difference.
This is because the LL result only depends on two-particle scattering, and
although this action could change the sign of certain amplitudes, it does
not change the collision rate. For example, the amplitudes of $%
q_{a}q_{b}\rightarrow $ $q_{a}q_{b}$ and $\bar{q}_{a}q_{b}\rightarrow $ $%
\bar{q}_{a}q_{b}$ ($a\neq b$) have different signs because one of the
couplings changes sign when we change the color into its anti-color, but the
amplitude squared is of the same. This makes the diagonal terms even in all
the chemical potentials 
\begin{equation}
\lambda _{aa}\left( \pm \mu _{1},\pm \mu _{2},\cdots ,\pm \mu
_{N_{f}}\right) =\lambda _{aa}\left( \mu _{1},\mu _{2},\cdots ,\mu
_{N_{f}}\right) ,
\end{equation}%
while the off-diagonal term $\lambda _{ab}$ is odd in $\mu _{a}$ and $\mu
_{b}$ but even in other chemical potentials 
\begin{align}
& \lambda _{ab}\left( \pm \mu _{1},\cdots ,\pm \mu _{a},\cdots ,\pm \mu
_{b},\cdots ,\pm \mu _{N_{f}}\right)  \notag \\
& =\text{sign}\left( \mu _{a}\mu _{b}\right) \lambda _{ab}\left( \mu
_{1},\cdots ,\mp \mu _{a},\cdots ,\mp \mu _{b},\cdots ,\mu _{N_{f}}\right) .
\label{lab}
\end{align}%
Thus, at the LL order, $\lambda $ becomes diagonal when all the chemical
potentials vanish.

To understand the other features of $\tilde{\lambda}_{qq}$ and $\tilde{%
\lambda}_{qq^{\prime }}$, we first turn to $\tilde{\lambda}_{+}$ and $\tilde{%
\lambda}_{-}$ in the $(T/\mu )^{2}\rightarrow 0$ limit. In this large
chemical potential limit, the quark contribution dominates over those of
anti-quark and gluon. The Fermi-Dirac distribution function $f^{q_{a}}$ of
quark $q_{a}$ multiplied by its Pauli blocking factor $F^{q_{a}}$ can be
well approximated by a $\delta -$function, $f_{{}}^{q_{a}}F_{{}}^{q_{a}}%
\simeq T\delta (E_{p}-\mu )$.

We then first set $\bm{\nabla }(\beta \mu _{a})=\bm{\nabla }(\beta \mu )$
for all $a$\ so all the currents $\bm{\nu }_{a}$ becomes identical. $\lambda
_{+}$ can be rewritten as $\sum_{a,b}\lambda _{ab}$, and Eq. (\ref{01})
yields%
\begin{equation}
\lambda _{+}\simeq -\frac{\beta }{3}N_{q}\int \frac{d^{3}p}{(2\pi )^{3}}%
T\delta (E_{p}-\mu )\mathbf{p}\cdot \mathbf{A}\sum_{a,b,c=1}^{N_{f}}\left( 
\frac{n_{b}}{\epsilon +P}-\frac{1}{\mu }\delta _{ca}\right) .
\end{equation}%
The summation gives $N_{f}^{2}\left( \frac{n\mu }{\epsilon +P}-1\right)
\propto N_{f}^{2}Ts/(\epsilon +P)\propto N_{f}^{2}T^{2}/\mu ^{2}$ and $%
\lambda _{+}\propto N_{f}^{2}T^{2}A$. On the other hand, Eq.(\ref{f1}) gives 
$\lambda _{+}\propto N_{f}^{4}\mu ^{2}A^{2}$ where $N_{f}^{4}$ comes from
summing the $a$, $b$, $c$, $d$ indices of $D_{q_{c}q_{d}\rightarrow
q_{c}q_{d}}^{ab}$ and we have used $%
f_{k_{1}}^{c_{1}}f_{k_{2}}^{c_{2}}F_{k_{3}}^{c_{3}}F_{k_{4}}^{c_{4}}\propto
T^{2}$ in Eq.(\ref{f2}). These two conditions yield $\lambda _{+}\propto
N_{f}^{0}T^{4}/\mu ^{2}$. This is indeed what happens in Fig. \ref{fig. 2}
at large $\mu $ (although the $1/\mu ^{2}$ dependence is not so obvious in
this plot but we have checked this at much larger $\mu ^{2}/T^{2}$).

We can perform the similar counting to the scaling of $\lambda _{-}$. From
Eq. (\ref{01}), $\lambda _{-}\propto N_{f}\mu ^{2}A$ and from Eq.(\ref{f1}) $%
\lambda _{-}\propto N_{f}^{3}\mu ^{2}A^{2}$. Thus, $\lambda _{-}\propto \mu
^{2}/N_{f}$ which is also observed in Fig. \ref{fig. 2}. The main difference
in $\lambda _{+}/N_{f}$ and $\lambda _{-}$ is the $T/\mu $ dependence---$%
\lambda _{-}$ has no cancellation factor of $\left( \frac{n\mu }{\epsilon +P}%
-1\right) \propto T^{2}$ in large $\mu $.

The different $\mu $ scaling between $\lambda _{+}$ and $\lambda _{-}$ at
large $\mu $ is due to collisions, which change the direction of the current
and reduce the conductivity. While both flavor singlet and non-singlet
fermions can collide among themselves, they do not collide with each other
(the scattering amplitude vanishes). Thus, when $\mu $, the flavor singlet
chemical potential, is increased, the flavor singlet current experiences
more collisions. Therefore the flavor singlet conductivity $\lambda _{+}$ is
reduced. For the flavor non-singlet current, the increase of $\mu $ does not
affect the collision. However, it will increase the averaged fermi momentum
such that the induced current and the flavor non-singlet conductivity $%
\lambda _{-}$ will be increased.

Given the large $\mu $ behavior of $\tilde{\lambda}_{+}$ and $\tilde{\lambda}%
_{-}$, the large $\mu $ behavior of $\tilde{\lambda}_{qq}$ and $\tilde{%
\lambda}_{qq^{\prime }}$ is now easily reconstructed: $\tilde{\lambda}%
_{qq}\simeq (N_{f}-1)\tilde{\lambda}_{-}/N_{f}\propto (N_{f}-1)\mu
^{2}/T^{2}N_{f}^{2}$ ($N_{f}\geq 2$) and $\tilde{\lambda}_{qq^{\prime
}}\simeq -\tilde{\lambda}_{-}/N_{f}\propto -\mu ^{2}/T^{2}N_{f}^{2}$. The
sign of $\tilde{\lambda}_{qq^{\prime }}$ can be best understood from the
flavor non-singlet current effect such that a gradient of $\mu _{a}$ induces
anti-$b$ currents ($b\neq a$) and yields $\tilde{\lambda}_{qq^{\prime }}<0$.
We can then interpolate $\tilde{\lambda}_{qq^{\prime }}$ to $\tilde{\lambda}%
_{qq^{\prime }}=0$ at zero $\mu $. There is no non-trivial structure at
intermediate $\mu $. For $\tilde{\lambda}_{qq}$, the $N_{f}=1$ curve seems
to be at odd with other $N_{f}$ curves, but this anomaly disappears when
viewed in the $\tilde{\lambda}_{+}$ plot.

The fact that $\lambda _{qq}>0$ while $\lambda _{qq^{\prime }}<0$ at finite $%
\mu $ is intriguing. It means a gradient $\bm{\nabla }\mu _{a}$ can drive a
current $\bm{\nu }_{a}$ along the $-\bm{\nabla }\mu _{a}$ direction, but it
will also drive currents of different flavors in the opposite direction.
This backward current phenomenon seems counter intuitive at the first sight.
But the physics behind is just that the flavor singlet current experiences
more collisions in a flavor singlet medium than the flavor non-singlet ones.
If the medium is flavor non-singlet, e.g. $\mu _{1}=-\mu _{2}\neq 0$ while
the other chemical potentials all vanish, then the flavor non-singlet
current $\bm{\nu }_{2}-\bm{\nu }_{1}$ will experience more collisions than
the flavor singlet current. Therefore, we will have $\lambda _{12}>0$. This
is consistent with Eq.(\ref{lab}) derived from the symmetry of the LL order
along. Thus the simple explanation based on collisions that we presented
above seems quite generic. It might happen in other systems such as cold
atoms as well. In that case, cold atom experiments might be the most
promising ones to observe this backward current phenomenon.

\begin{table}[tbp]
\caption{The coefficients in the $(\protect\mu /T)^{2}$ expansions of $%
\tilde{\protect\lambda}_{qq}$ and $\tilde{\protect\lambda}_{qq^{\prime }}$
for small $\protect\mu /T$. Our result for $\protect\mu /T=0$ agrees within $%
0.1\%$ to that of Arnold, Moore and Yaffe calculated up to $N_{f}=6$ 
\protect\cite{Arnold:2000dr}. }
\label{power expansion coefficients lambdaqq}\centering{}$%
\begin{tabular}{|c|r|r|c||c|c|c|c|}
\hline
$N_{f}$ & $a_{0}$\;\;\  & $a_{1}$\;\quad\  & $a_{1}^{\prime}$ & $N_{f}$ & $%
a_{0}$ & $a_{1}$ & $a_{1}^{\prime}$ \\ \hline
1 & 14.3676 & -0.3077 & - & 9 & 7.8019 & 2.1076 & -0.7572 \\ \hline
2 & 12.9989 & 1.7347 & -5.0372 & 10 & 7.3806 & 1.9880 & -0.6404 \\ \hline
3 & 11.8688 & 2.3969 & -3.3569 & 11 & 7.0025 & 1.8766 & -0.5487 \\ \hline
4 & 10.9197 & 2.5757 & -2.3922 & 12 & 6.6612 & 1.7731 & -0.4754 \\ \hline
5 & 10.1113 & 2.5680 & -1.7906 & 13 & 6.3517 & 1.6791 & -0.4159 \\ \hline
6 & 9.4145 & 2.4791 & -1.3909 & 14 & 6.0697 & 1.5917 & -0.3668 \\ \hline
7 & 8.8076 & 2.3600 & -1.1117 & 15 & 5.8117 & 1.5121 & -0.3260 \\ \hline
8 & 8.2743 & 2.2319 & -0.9090 & 16 & 5.5747 & 1.4384 & -0.2916 \\ \hline
\end{tabular}
\ $%
\end{table}

\section{Summary \label{sec:Conclusion-and-discussion}}

We have calculated the conductivity matrix of a weakly coupled quark-gluon
plasma at the leading-log order. By setting all quark chemical potentials to
be identical, the diagonal conductivities become degenerate and positive,
while the off-diagonal ones become degenerate but negative (or zero when the
chemical potential vanishes). This means a potential gradient of a certain
fermion flavor can drive backward currents of other flavors. A simple
explanation is provided for this seemingly counter intuitive phenomenon. It
is speculated that this phenomenon is generic and most easily measured in
cold atom experiments.

Acknowledgement: SP thanks Tomoi Koide and Xu-guang Huang for helpful
discussions on the Onsager relation. JWC thanks Jan M. Pawlowski for useful
discussions and the U. of Heidelberg for hospitality. JWC, YFL and SP are
supported by the CTS and CASTS of NTU and the NSC (102-2112-M-002-013-MY3)
of ROC. YKS is supported in part by the CCNU-QLPL Innovation Fund under
grant No. QLPL2011P01. This work is also supported by the National Natural
Science Foundation of China under grant No. 11125524 and 11205150, and in
part by the China Postdoctoral Science Foundation under the grant No.
2011M501046.


\begin{thebibliography}{99}
\bibitem{Arsene:2004fa} I.~Arsene \textit{et al.} [BRAHMS Collaboration], 
%``Quark gluon plasma and color glass condensate at RHIC? The Perspective from the BRAHMS experiment,''
Nucl.\ Phys.\ A \textbf{757}, 1 (2005) [nucl-ex/0410020]. 
%%CITATION = NUCL-EX/0410020;%%

\bibitem{Adcox:2004mh} K.~Adcox \textit{et al.} [PHENIX Collaboration], 
%``Formation of dense partonic matter in relativistic nucleus-nucleus collisions at RHIC: Experimental evaluation by the PHENIX collaboration,''
Nucl.\ Phys.\ A \textbf{757}, 184 (2005) [nucl-ex/0410003]. 
%%CITATION = NUCL-EX/0410003;%%

\bibitem{Back:2004je} B.~B.~Back, M.~D.~Baker, M.~Ballintijn, D.~S.~Barton,
B.~Becker, R.~R.~Betts, A.~A.~Bickley and R.~Bindel \textit{et al.}, 
%``The PHOBOS perspective on discoveries at RHIC,''
Nucl.\ Phys.\ A \textbf{757}, 28 (2005) [nucl-ex/0410022]. 
%%CITATION = NUCL-EX/0410022;%%

\bibitem{Adams:2005dq} J.~Adams \textit{et al.} [STAR Collaboration], 
%``Experimental and theoretical challenges in the search for the quark gluon plasma: The STAR Collaboration's critical assessment of the evidence from RHIC collisions,''
Nucl.\ Phys.\ A \textbf{757}, 102 (2005) [nucl-ex/0501009]. 
%%CITATION = NUCL-EX/0501009;%%

\bibitem{Song:2010mg} H.~Song, S.~A.~Bass, U.~Heinz, T.~Hirano and C.~Shen, 
%``200 A GeV Au+Au collisions serve a nearly perfect quark-gluon liquid,''
Phys.\ Rev.\ Lett.\ \textbf{106}, 192301 (2011) [Erratum-ibid.\ \textbf{109}%
, 139904 (2012)]. %%CITATION = ARXIV:1011.2783;%%

\bibitem{Kovtun:2004de} P.~Kovtun, D.~T.~Son and A.~O.~Starinets, Phys.\
Rev.\ Lett.\ \textbf{94}, 111601 (2005). 
%``Viscosity in strongly interacting quantum field theories from black hole %physics,''
%%CITATION = PRLTA,94,111601;%%

\bibitem{Maldacena:1997re} J.~M.~Maldacena, Adv.\ Theor.\ Math.\ Phys.\ 
\textbf{2}, 231 (1998) [Int.\ J.\ Theor.\ Phys.\ \textbf{38}, 1113 (1999)]. 
%[arXiv:hep-th/9711200].
%%CITATION = IJTPB,38,1113;%

\bibitem{Gubser:1998bc} S.~S.~Gubser, I.~R.~Klebanov and A.~M.~Polyakov,
Phys.\ Lett.\ B \textbf{428}, 105 (1998) 
%``Gauge theory correlators from non-critical string theory,''
%%CITATION = PHLTA,B428,105;%%

\bibitem{Witten:1998qj} E.~Witten, Adv.\ Theor.\ Math.\ Phys.\ \textbf{2},
253 (1998) %[arXiv:hep-th/9802150].
%``Anti-de Sitter space and holography,''
%%CITATION = 00203,2,253;%%

\bibitem{Arnold:2000dr} P.~B.~Arnold, G.~D.~Moore and L.~G.~Yaffe, 
%``Transport coefficients in high temperature gauge theories. 1. Leading log results,''
JHEP \textbf{0011}, 001 (2000) [hep-ph/0010177]. 
%%CITATION = HEP-PH/0010177;%%
%\cite{Jeon:1994if}
%\cite{Arnold:2006fz}

\bibitem{Arnold:2003zc} P.~B.~Arnold, G.~D. Moore and L.~G.~Yaffe, 
%``Transport coefficients in high temperature gauge theories. 2. Beyond leading log,''
JHEP \textbf{0305}, 051 (2003) [hep-ph/0302165]. 
%%CITATION = HEP-PH/0302165;%%

\bibitem{Chen:2010xk} J.-W.~Chen, J.~Deng, H.~Dong and Q.~Wang, 
%``How Perfect a Gluon Plasma Can Be in Perturbative QCD?,''
Phys.\ Rev.\ D \textbf{83}, 034031 (2011) [Erratum-ibid.\ D \textbf{84},
039902 (2011)] [arXiv:1011.4123 [hep-ph]].

%\cite{Chen:2011km}

\bibitem{Chen:2011km} J.~-W.~Chen, J.~Deng, H.~Dong and Q.~Wang, 
%``Shear and Bulk Viscosities of a Gluon Plasma in Perturbative QCD: Comparison of Different Treatments for the gg<->ggg Process,''
Phys.\ Rev.\ C \textbf{87}, 024910 (2013) [arXiv:1107.0522 [hep-ph]]. 
%%CITATION = ARXIV:1107.0522;%%
%6 citations counted in INSPIRE as of 11 Aug 2013

\bibitem{Csernai:2006zz} L.~P.~Csernai, J.~I.~Kapusta and L.~D.~McLerran, 
%``On the strongly-interacting low-viscosity matter created in relativistic
%nuclear collisions,''
Phys.\ Rev.\ Lett.\ \textbf{97}, 152303 (2006).

\bibitem{Chen:2006iga} J.-W.~Chen and E.~Nakano, 
%``Shear viscosity to entropy density ratio of QCD below the deconfinement temperature,''
Phys.\ Lett.\ B \textbf{647}, 371 (2007) [hep-ph/0604138]. 
%%CITATION = HEP-PH/0604138;%%

%\cite{Chen:2012jc}

\bibitem{Chen:2012jc} J.-W.~Chen, Y.~-F.~Liu, Y.-K.~Song and Q.~Wang, 
%``Shear and bulk viscosities of a weakly coupled quark gluon plasma with finite chemical potential and temperature: Leading-log results,''
Phys.\ Rev.\ D \textbf{87}, 036002 (2013) [arXiv:1212.5308 [hep-ph]]. 
%%CITATION = ARXIV:1212.5308;%%
%1 citations counted in INSPIRE as of 07 Aug 2013

\bibitem{Chen:2007xe} J.-W.~Chen, Y.~-H.~Li, Y.~-F.~Liu and E.~Nakano, 
%``QCD viscosity to entropy density ratio in the hadronic phase,''
Phys.\ Rev.\ D \textbf{76}, 114011 (2007) [hep-ph/0703230]. 
%%CITATION = HEP-PH/0703230;%%

\bibitem{Kharzeev:2007wb} D.~Kharzeev, K.~Tuchin, 
%``Bulk viscosity of QCD matter near the critical temperature,''
JHEP \textbf{0809}, 093 (2008).

\bibitem{Karsch:2007jc} F.~Karsch, D.~Kharzeev, K.~Tuchin, 
%``Universal properties of bulk viscosity near the QCD phase transition,''
Phys.\ Lett.\ \textbf{B663}, 217-221

\bibitem{Meyer:2010ii} H.~B.~Meyer, 
%``The Bulk Channel in Thermal Gauge Theories,''
JHEP \textbf{1004}, 099 (2010). %%CITATION = JHEPA,1004,099;%%

\bibitem{Arnold:2006fz} P.~B.~Arnold, C.~Dogan and G.~D.~Moore, 
%``The Bulk Viscosity of High-Temperature QCD,''
Phys.\ Rev.\ D \textbf{74}, 085021 (2006) [hep-ph/0608012]. 
%%CITATION = HEP-PH/0608012;%%

\bibitem{Chen:2007kx} J.-W.~Chen and J.~Wang, 
%``Bulk viscosity of a gas of massless pions,''
Phys.\ Rev.\ C \textbf{79}, 044913 (2009) [arXiv:0711.4824 [hep-ph]]. 
%%CITATION = ARXIV:0711.4824;%%

\bibitem{FernandezFraile:2008vu} D.~Fernandez-Fraile and A.~Gomez Nicola, 
%``Bulk viscosity and the conformal anomaly in the pion gas,''
Phys.\ Rev.\ Lett.\ \textbf{102}, 121601 (2009). 
%%CITATION = ARXIV:0809.4663;%%

\bibitem{Lu:2011df} E.~Lu, G.~D.~Moore, 
%``The Bulk Viscosity of a Pion Gas,''
Phys.\ Rev.\ \textbf{C83}, 044901 (2011).

\bibitem{Dobado:2011qu} A.~Dobado, F.~J.~Llanes-Estrada and
J.~M.~Torres-Rincon, 
%``Bulk viscosity of low-temperature strongly interacting matter,''
Phys.\ Lett.\ B \textbf{702}, 43 (2011). %%CITATION = PHLTA,B702,43;%%

%\cite{Chakraborty:2010fr}

\bibitem{Chakraborty:2010fr} P.~Chakraborty and J.~I.~Kapusta, 
%``Quasi-Particle Theory of Shear and Bulk Viscosities of Hadronic Matter,''
Phys.\ Rev.\ C \textbf{83}, 014906 (2011). %[arXiv:1006.0257 [nucl-th]].
%%CITATION = ARXIV:1006.0257;%%

%\cite{Dong:2007mb}

\bibitem{Dong:2007mb} H.~Dong, N.~Su and Q.~Wang, 
%``Baryon number conservation and enforced electric charge neutrality for bulk viscosity in quark matter,''
Phys.\ Rev.\ D \textbf{75}, 074016 (2007) [astro-ph/0702104]. 
%%CITATION = ASTRO-PH/0702104;%%
%13 citations counted in INSPIRE as of 10 Aug 2013

%\cite{Alford:2006gy}

\bibitem{Alford:2006gy} M.~G.~Alford and A.~Schmitt, 
%``Bulk viscosity in 2SC quark matter,''
J.\ Phys.\ G \textbf{34}, 67 (2007) [nucl-th/0608019]. 
%%CITATION = NUCL-TH/0608019;%%
%39 citations counted in INSPIRE as of 10 Aug 2013

%\cite{Alford:2008pb}

\bibitem{Alford:2008pb} M.~G.~Alford, M.~Braby and A.~Schmitt, 
%``Bulk viscosity in kaon-condensed color-flavor locked quark matter,''
J.\ Phys.\ G \textbf{35}, 115007 (2008) [arXiv:0806.0285 [nucl-th]]. 
%%CITATION = ARXIV:0806.0285;%%
%23 citations counted in INSPIRE as of 10 Aug 2013  

%\cite{Sa'd:2006qv}

\bibitem{Sa'd:2006qv} B.~A.~Sa'd, I.~A.~Shovkovy and D.~H.~Rischke, 
%``Bulk viscosity of spin-one color superconductors with two quark flavors,''
Phys.\ Rev.\ D \textbf{75}, 065016 (2007) [astro-ph/0607643]. 
%%CITATION = ASTRO-PH/0607643;%%
%40 citations counted in INSPIRE as of 10 Aug 2013  

%\cite{Sa'd:2007ud}

\bibitem{Sa'd:2007ud} B.~A.~Sa'd, I.~A.~Shovkovy and D.~H.~Rischke, 
%``Bulk viscosity of strange quark matter: Urca versus non-leptonic processes,''
Phys.\ Rev.\ D \textbf{75}, 125004 (2007) [astro-ph/0703016]. 
%%CITATION = ASTRO-PH/0703016;%%
%30 citations counted in INSPIRE as of 10 Aug 2013

%\cite{Wang:2010ydb}

\bibitem{Wang:2010ydb} X.~Wang and I.~A.~Shovkovy, 
%``Bulk viscosity of spin-one color superconducting strange quark matter,''
Phys.\ Rev.\ D \textbf{82}, 085007 (2010) [arXiv:1006.1293 [hep-ph]]. 
%%CITATION = ARXIV:1006.1293;%%
%6 citations counted in INSPIRE as of 10 Aug 2013

%\cite{Israel:1979wp}

\bibitem{Huang:2013iia} X.~-G.~Huang and J.~Liao, 
%``Axial Current Generation from Electric Field: Chiral Electric Separation Effect,''
Phys.\ Rev.\ Lett.\ \textbf{110}, 232302 (2013) [arXiv:1303.7192 [nucl-th]]. 
%%CITATION = ARXIV:1303.7192;%%
%3 citations counted in INSPIRE as of 12 Aug 2013

%\cite{McLerran:2013hla}

\bibitem{McLerran:2013hla} L.~McLerran and V.~Skokov, 
%``Comments About the Electromagnetic Field in Heavy-Ion Collisions,''
arXiv:1305.0774 [hep-ph]. %%CITATION = ARXIV:1305.0774;%%
%5 citations counted in INSPIRE as of 12 Aug 2013

%\cite{Ding:2010ga}

\bibitem{Ding:2010ga} H.~-T.~Ding, A.~Francis, O.~Kaczmarek, F.~Karsch,
E.~Laermann and W.~Soeldner, 
%``Thermal dilepton rate and electrical conductivity: An analysis of vector current correlation functions in quenched lattice QCD,''
Phys.\ Rev.\ D \textbf{83}, 034504 (2011) [arXiv:1012.4963 [hep-lat]]. 
%%CITATION = ARXIV:1012.4963;%%
%54 citations counted in INSPIRE as of 12 Aug 2013

%\cite{Amato:2013naa}

\bibitem{Amato:2013naa} A.~Amato, G.~Aarts, C.~Allton, P.~Giudice, S.~Hands
and J.~-I.~Skullerud, 
%``Electrical conductivity of the quark-gluon plasma across the deconfinement transition,''
arXiv:1307.6763 [hep-lat]. %%CITATION = ARXIV:1307.6763;%%

%\cite{Qin:2013aaa}

\bibitem{Qin:2013aaa} S.~-x.~Qin, 
%``A Divergence-Free Method to Extract Observables from Meson Correlation Functions,''
arXiv:1307.4587 [nucl-th]. %%CITATION = ARXIV:1307.4587;%%
%1 citations counted in INSPIRE as of 12 Aug 2013

\bibitem{Israel:1979wp} W.~Israel and J.~M.~Stewart, 
%``Transient relativistic thermodynamics and kinetic theory,''
Annals Phys.\ \textbf{118}, 341 (1979). %%CITATION = APNYA,118,341;%%
%491 citations counted in INSPIRE as of 08 Aug 2013

%\cite{Pu:2011vr}

\bibitem{Pu:2011vr} S.~Pu, 
%``Relativistic fluid dynamics in heavy ion collisions,''
arXiv:1108.5828 [hep-ph]. %%CITATION = ARXIV:1108.5828;%%

\bibitem{Jeon:1994if} S.~Jeon, 
%``Hydrodynamic transport coefficients in relativistic scalar field theory,''
Phys.\ Rev.\ D \textbf{52}, 3591 (1995). %%CITATION = PHRVA,D52,3591;%%

%\cite{Jeon:1995zm}

\bibitem{Hidaka:2010gh} Y.~Hidaka and T.~Kunihiro, 
%``Renormalized Linear Kinetic Theory as Derived from Quantum Field Theory: A Novel diagrammatic method for computing transport coefficients,''
Phys.\ Rev.\ D \textbf{83}, 076004 (2011).

\bibitem{Gagnon:2007qt} J.~S.~Gagnon and S.~Jeon, 
%``Leading Order Calculation of Shear Viscosity in Hot Quantum Electrodynamics
%from Diagrammatic Methods,''
Phys.\ Rev.\ D \textbf{76}, 105019 (2007).

\bibitem{Arnold:2002zm} P.~B.~Arnold, G.~D.~Moore and L.~G.~Yaffe, 
%``Effective kinetic theory for high temperature gauge theories,''
JHEP \textbf{0301}, 030 (2003) [hep-ph/0209353]. 
%%CITATION = HEP-PH/0209353;%%
%\cite{Mrowczynski:2000ed}

\bibitem{Mrowczynski:2000ed} S.~Mrowczynski and M.~H.~Thoma, 
%``Hard loop approach to anisotropic systems,''
Phys.\ Rev.\ D \textbf{62}, 036011 (2000) [hep-ph/0001164]. 
%%CITATION = HEP-PH/0001164;%%

%\cite{Huang:2013iia}
\end{thebibliography}
\end{document}